\documentclass[prl,twocolumn,nofootinbib,floatfix,subeqn]{revtex4-1}
\usepackage[usenames]{color}
\usepackage{amsmath}
\usepackage{amssymb}
\usepackage{float}
\usepackage{graphicx}
\usepackage{xspace}
\usepackage{hyperref}
\usepackage{hhline}
\usepackage{bbold}

\newcommand{\ket}[1]{\ensuremath{\left|#1\right\rangle}}

\newcommand{\OP}{{\cal O}_P}

\newcommand{\avs}[1]{\langle  #1  \rangle}

\long\def\symbolfootnote[#1]#2{\begingroup%
\def\thefootnote{\fnsymbol{footnote}}\footnotetext[#1]{#2}\endgroup}

\begin{document}

\title{Observation of Correlated Particle-Hole Pairs and\\String Order in Low-Dimensional Mott Insulators}


\author{M. Endres$^{1,*}$}
\author{M. Cheneau$^{1}$}
\author{T. Fukuhara$^{1}$}
\author{C. Weitenberg$^{1}$}
\author{P. Schau\ss$^{1}$}
\author{C.~Gross$^{1}$}
\author{L. Mazza$^{1}$}
\author{M.C. Ba{\~n}uls$^1$}
\author{L. Pollet$^{2}$}
\author{I. Bloch$^{1,3}$}
\author{S. Kuhr$^{1,4}$}

\date{16 August 2011}

\affiliation{\vspace{0.2cm}$^1$Max-Planck-Institut f\"ur Quantenoptik, 85748 Garching, Germany}
\affiliation{$^2$ETH Zurich, 8093 Zurich, Switzerland}
\affiliation{$^3$Ludwig-Maximilians-Universit\"at, Schellingstr.~4/II, 80799 M\"unchen} \affiliation{$^4$University of Strathclyde, SUPA, Glasgow G4 0NG, United Kingdom}

\begin{abstract}
Quantum phases of matter are characterized by the underlying correlations of the many-body system. Although this is typically captured by a local order parameter, it has been shown that a broad class of many-body systems possesses a hidden non-local order. In the case of bosonic Mott insulators, the ground state properties are governed by quantum fluctuations in the form of correlated particle-hole pairs that lead to the emergence of a non-local string order in one dimension. Using high-resolution imaging of low-dimensional quantum gases in an optical lattice, we directly detect these pairs  with single-site and single-particle sensitivity and observe string order in the one-dimensional case.
\end{abstract}

\maketitle

\symbolfootnote[1]{Electronic address: {\bf manuel.endres@mpq.mpg.de}}


The realization of strongly correlated quantum many-body systems using ultracold atoms has enabled the direct observation and control of fundamental quantum effects \cite{Jaksch:2005,Lewenstein:2007,Bloch:2008c}. A prominent example is the transition from a superfluid (SF) to a Mott insulator (MI), occurring when interactions between bosonic particles on a lattice dominate over their kinetic energy \cite{Fisher:1989,Jaksch:1998,Greiner:2002a,Stoeferle:2004,Spielman:2007}. At zero temperature, and in the limit where the ratio of kinetic energy over interaction energy vanishes, particle fluctuations are completely suppressed and the lattice sites are occupied by an integer number of particles. However, at a finite tunnel coupling, but still in the Mott insulating regime, quantum fluctuations create correlated particle-hole pairs on top of this fixed-density background, which can be understood as virtual excitations. These particle-hole pairs fundamentally determine the properties of the Mott insulator such as its residual phase coherence \cite{Gerbier:2005a} and lie at the heart of superexchange-mediated spin interactions that form the basis of quantum magnetism in multi-component quantum gas mixtures \cite{Kuklov:2003,Duan:2003,Trotzky:2008a}.

In a one-dimensional system, the appearance of correlated particle-hole pairs at the transition point from a superfluid to a Mott insulator is intimately connected to the emergence of a hidden string-order parameter ${\cal O}_P$ \cite{DallaTorre:2006,Berg:2008}:
\begin{equation}
	{\cal O}_P^2 = \lim_{l \rightarrow \infty} {\cal O}_P^2(l) = \lim_{l \rightarrow \infty} \left\langle \prod_{k\leq j \leq k+l}e^{i \pi \delta \hat n_j}\right \rangle.\label{eq:stringorder}
\end{equation}
Here $\delta \hat n_j = \hat n_j - \bar n$ denotes the deviation in occupation of the $j$th lattice site from the average background density, and $k$ is an arbitrary position along the chain. In the simplest case of a Mott insulator with unity filling ($\bar n = 1$), relevant to our experiments, each factor in the product of operators in Eq.\,\ref{eq:stringorder} yields $-1$ instead of $+1$, when a single-particle fluctuation from the unit background density is encountered. In the superfluid, particle and hole fluctuations occur independently and are uncorrelated, such that $\OP=0$. However, in the Mott insulating phase, density fluctuations always occur as correlated particle-hole pairs, resulting in $\OP \neq 0$. For a homogeneous system, $\OP$ is expected to follow a scaling of Berezinskii-Kosterlitz-Thouless (BKT) type \cite{Kuehner:1998}. Non-local correlation functions, like the string-order parameter defined above, have been introduced in the context of low-dimensional quantum systems. They classify many-body quantum phases that are not amenable to a description through a local order parameter, typically used in the Landau paradigm of phase transitions. Examples include spin-1 chains \cite{denNijs:1989} and spin-1/2 ladders \cite{Kim:2000}, fermionic Mott and band insulators \cite{Anfuso:2007b} and Haldane insulators in one-dimensional Bose gases \cite{DallaTorre:2006,Berg:2008}. Recently, the intimate connection of string order and local symmetries has been uncovered \cite{PerezGarcia:2008} and wide-ranging classification schemes for quantum phases using such symmetry principles have been introduced \cite{Chen:2011,Schuch:2010}. Here we show for the first time that  correlated particle-hole pairs and string order can be directly detected using single-atom-resolved images of strongly correlated ultracold quantum gases \cite{Bakr:2010,Sherson:2010}.

\begin{figure}
\begin{center}
\includegraphics[width=0.8\columnwidth]{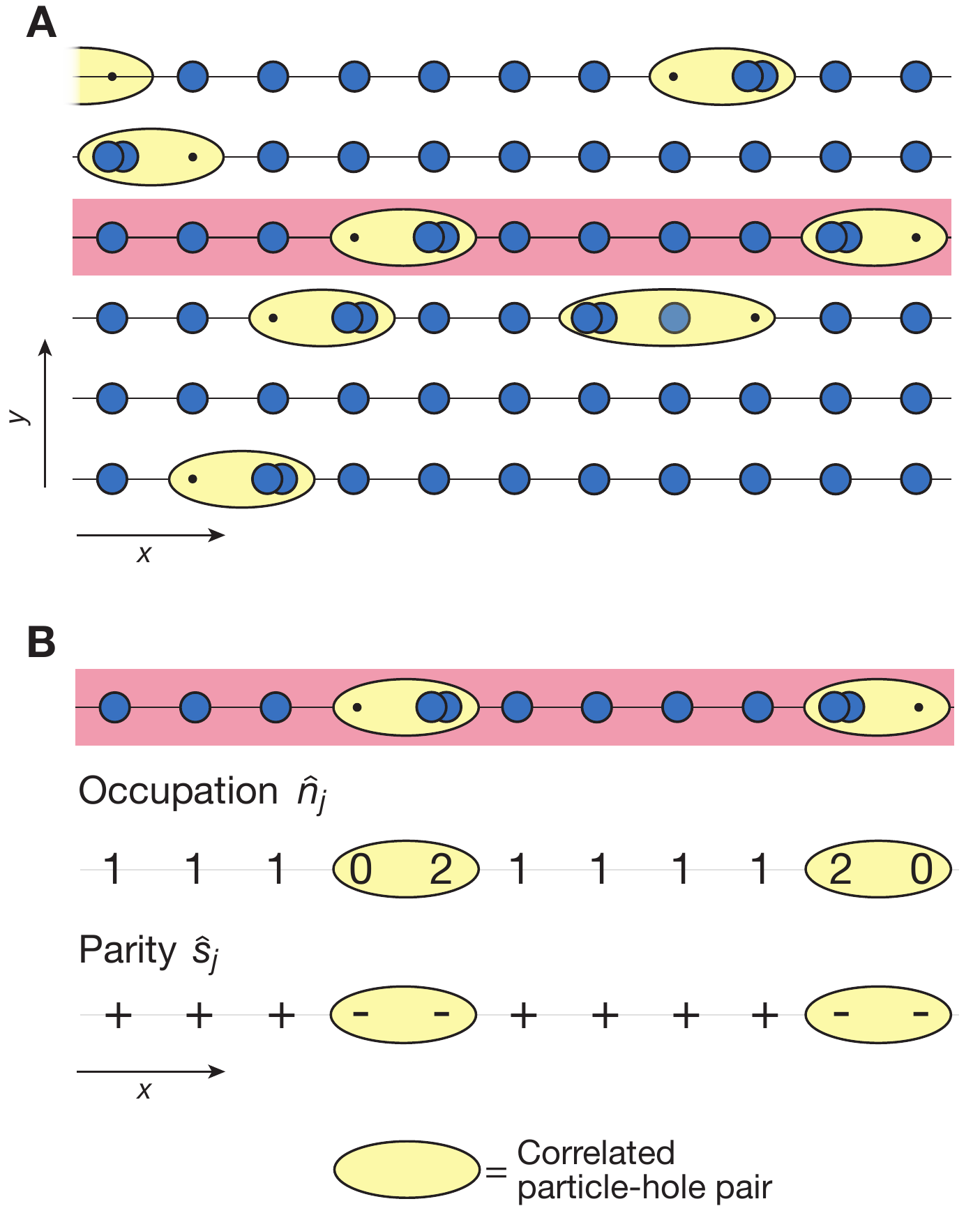}
\vspace{-0.5cm}
\end{center}
\caption{ {\bf Quantum-correlated particle-hole pairs in one-dimensional Mott insulators.} ({\bf A}) In a two-dimensional array of atoms (blue circles), decoupled one-dimensional systems were created by suppressing tunneling along the $y$ direction. Quantum fluctuations then only induce correlated particle-hole excitations (yellow ellipses) along the $x$ direction of the one-dimensional Mott insulators. ({\bf B}) Such correlated fluctuations in the occupation $\hat n_j$ are detected in the experiment as correlated fluctuations in the parity $\hat s_j$. The light red bar in (A) marks the one-dimensional chain chosen in (B) for further explanations. \label{fig:Schematic}}
\end{figure}

We prepared a two-dimensional degenerate gas of ultracold $^{87}$Rb atoms, before shining in a two-dimensional square optical lattice (lattice spacing $a_{\rm lat}=532$\,nm)  with variable lattice depths in $x$ and $y$ directions \cite{Sherson:2010} (see Appendix). A microscope objective with a resolution comparable to the lattice spacing was used for fluorescence detection of individual atoms. Because inelastic light-assisted collisions during the imaging lead to a rapid loss of  atom pairs, our scheme detects the parity of the atom number. We used an algorithm to deconvolve the images, yielding single-site-resolved information of the on-site parity. Typically, our samples  contained  $150 - 200$ atoms in order to avoid  MIs of occupation numbers $\bar n >1$.

To detect particle-hole pairs, we evaluated two-site parity correlation functions \cite{Kapit:2010}
\begin{equation}
	C(d) = \avs{\hat s_k \hat s_{k+d}} - \avs{\hat s_k}\avs{\hat s_{k+d}}\label{eq:Cd},
\end{equation}
where $\hat s_k=e^{i \pi \delta \hat n_k}$ is the parity operator at site $k$ and $d$ is the distance between the lattice sites.  For the case of $\bar n=1$, $\hat s_k$ yields $+1(-1)$  for an odd(even) occupation number $n_k$. If a particle-hole pair exists on sites $k$ and $k+d$, the same parity $s(n_k)=s(n_{k+d})=-1$ is detected (Fig.\,\ref{fig:Schematic}). The existence of correlated particle-hole pairs therefore leads to an increase of $ \avs{\hat s_k \hat s_{k+d}}$ above the factorized form $\avs{\hat s_k}\avs{\hat s_{k+d}}$, which results from uncorrelated fluctuations, e.g., due to thermal excitations. We obtained $C(d)$  from our deconvolved images by an average over many experimental realizations and by an additional average over $k$ in a central region of interest.

\begin{figure}[!b]
\begin{center}
\includegraphics[width=0.92\columnwidth]{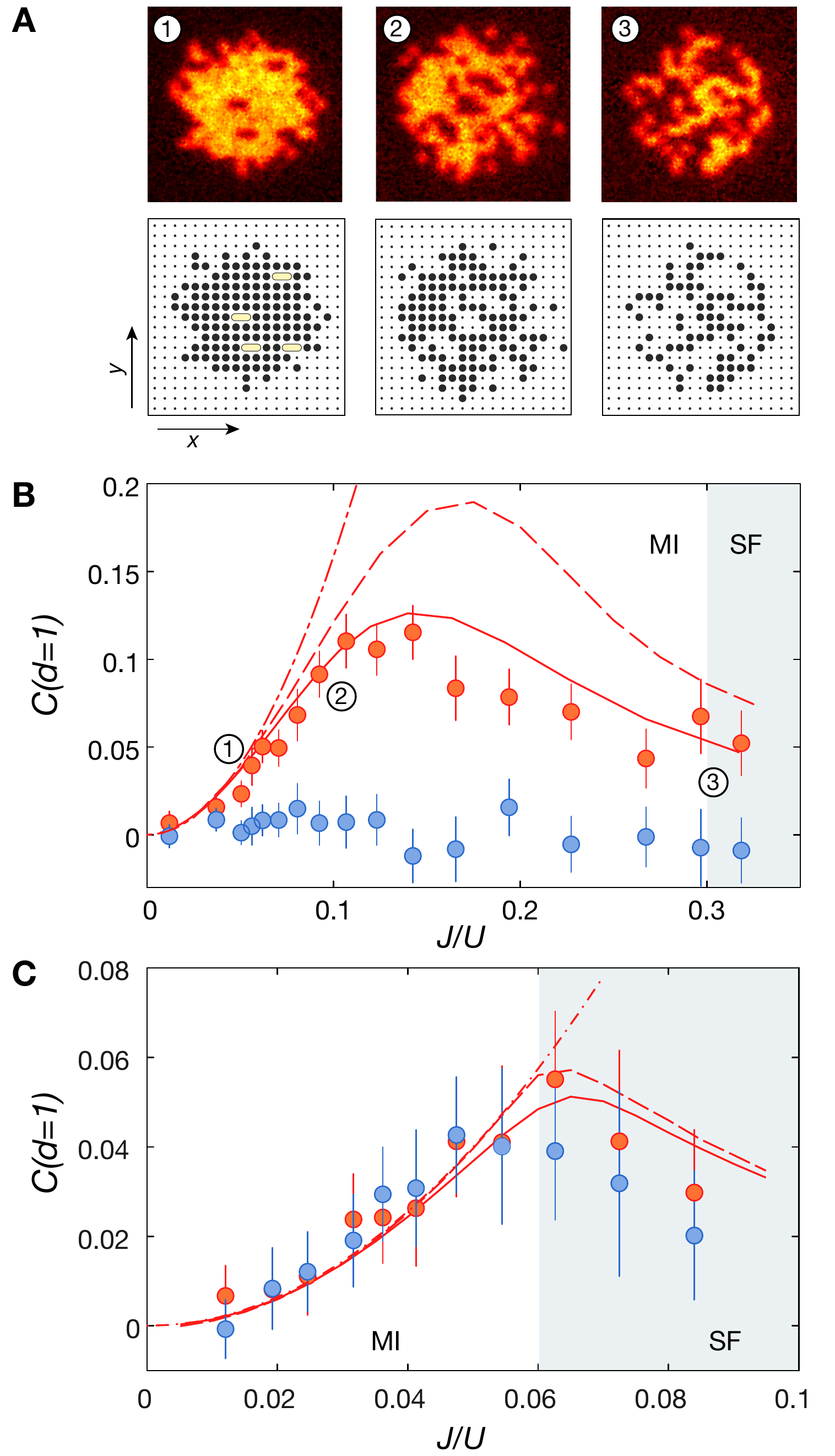}
\end{center}
\vspace{-0.4cm}
\caption{ {\bf Non-local parity correlations.} ({\bf A}) Top row: Typical experimental fluorescence images for $J/U=0.06$ (A1), $J/U=0.11$ (A2) and $J/U=0.3$ (A3) for the one-dimensional geometry.   Bottom row: Reconstructed on-site parity. Particle-hole pairs are emphasized by a yellow shading in (A1). For increased $J/U$ the pairs start to proliferate and an identification in a single experimental image becomes impossible (A2, A3)  ({\bf B}) One-dimensional nearest-neighbor correlations $C(d=1)$ as a function of $J/U$ along the $x$ (red circles) and $y$ directions (blue circles). The curves are first-order perturbation theory (dashed-dotted line),  Density-Matrix Renormalization Group (DMRG) calculations for a homogeneous system at $T=0$ (dashed line) and finite-temperature MPS calculations including harmonic confinement at $T=0.09\,U/k_B$ (solid line). ({\bf C}) Parity correlations in two dimensions. Symbols have the same meaning as in (B). The curves are first-order perturbation theory (dashed-dotted line) and a QMC calculation for a homogeneous system at $T=0.01\,U/k_B$ (dashed line) and $T=0.1\,U/k_B$ (solid line).  Each data point is an average over the central $9 \times 7$ lattice sites from $50-100$ pictures. The error bars denote the 1$\sigma$ statistical uncertainty. The light blue shading highlights the SF phase.}\label{fig:nncorrelation}
\end{figure}

We first analyzed two-site parity correlations in one-dimensional systems (Fig.\,\ref{fig:nncorrelation}A and B).
To create isolated one-dimensional tubes, we kept the lattice axis along $y$ at  a constant depth of $V_y=17(1)\,E_r$, where $E_{r}=h^2/(8m a_{\rm lat}^2)$ denotes the recoil energy and $m$ is the atomic mass of $^{87}$Rb. We recorded the nearest-neighbor correlations $C(d=1)$ for different values of $J/U$ along the direction of the one-dimensional tubes (red circles in Fig.\,\ref{fig:nncorrelation}B), where $J$ and $U$ are the tunneling matrix element and the on-site interaction energy in the Bose-Hubbard model, respectively  (see Appendix). For small $J/U$, the nearest-neighbor correlations vanish, since only uncorrelated thermal excitations exist deep in the MI regime. As particle-hole pairs emerge with increasing $J/U$, we observe an increase of nearest-neighbor correlations until a peak value is reached, well before the critical value $(J/U)_c^{1d}\approx 0.3$ \cite{Kuehner:2000,Kashurnikov:1996a} for the  one-dimensional SF-MI transition. The observed signal is a genuine quantum effect because thermally induced particle-hole pairs extend over arbitrary distances and are therefore uncorrelated. Their presence leads to a reduction of the correlation signal. We found no correlations when performing the same analysis perpendicular to the one-dimensional tubes (blue circles in Fig.\,\ref{fig:nncorrelation}B), showing that the coupling between the tubes was negligible.

Our data show very good agreement with ab-initio finite-temperature Matrix Product State (MPS) calculations \cite{Verstraete:2004,Zwolak:2004} at temperature $T=0.09\,U/k_B$  (Fig.\,\ref{fig:nncorrelation}A, solid line) that also take into account our harmonic trapping potential with frequency $\omega/(2\pi) = 60(1)\rm{Hz}$. Compared to a homogeneous system at $T=0$ (dashed-line), the experimental signal is reduced, especially around the maximum. This reduction can be attributed in equal parts to the finite temperature of our system and the averaging over different local chemical potentials. The latter is especially severe in the one-dimensional case owing to the narrow width of the Mott lobe for $\bar n=1$ close to the critical point \cite{Kuehner:1998}.  Interestingly, the growth of particle-hole correlations $\propto J^2/U^2$ expected from first-order perturbation theory  (see Appendix) is limited to very small values $J/U<0.05$, before significant deviations in the experiment and the numerical simulations are observed.

As the dimensionality of the system plays an important role in its correlation properties, we also measured the two-site parity correlations across the two-dimensional  SF-MI transition  by simultaneously varying $J/U$ along both lattice axes (Fig.\,\ref{fig:nncorrelation}C). In contrast to the one-dimensional case, we now clearly observe the same nearest-neighbor correlations within our error bars along both axes. The maximum correlations are significantly smaller than in one dimension, and the peak value is now reached around the critical value $(J/U)_c^{2d}\approx0.06$  \cite{Capogrosso:2008}. We compared our data with Quantum Monte Carlo (QMC) simulations for a homogeneous system at  $T=0.1\,U/k_B$ (solid line in Fig.\,\ref{fig:nncorrelation}C) and found good quantitative agreement. Here, the broader shape of the Mott lobe leads to a weaker averaging effect over different local chemical potentials. The increased strength of the correlations and the larger shift of the maximum of the correlations relative to the critical point in the one-dimensional case directly reflect the more prominent role of quantum fluctuations in lower dimensions. This can also be seen from the on-site fluctuations $C(d=0)$ at the critical point which are significantly increased in the one-dimensional case (see Appendix). In both the one-dimensional and the two-dimensional systems, two-site correlations are expected to decay strongly with distance. Our data for the next-nearest-neighbor correlations $C(d=2)$ is consistent with this predicted behavior  (see Appendix).

In addition to  two-site correlations, we evaluated string-type correlators ${\cal O}_P^2(l)= \left\langle \prod_{j=k}^{k+l} \hat{s}_j \right \rangle$ (see Eq.\,\ref{eq:stringorder} and Fig.\,\ref{fig:Schematic}B) in our one-dimensional systems, where the product is calculated over a chain of length $l+1$. In the simplest case of a zero-temperature MI at $J/U=0$, no fluctuations exist  and therefore ${\cal O}_P^2(l)=1$. As $J/U$ increases, fluctuations in the form of particle-hole pairs appear. Whenever a certain number of  particle-hole pairs lies completely within the region covered by the string correlator, the respective minus signs cancel pairwise. However, there is also the possibility that a particle-hole pair is cut by one end of the string correlator, for example when a particle exists at position $<k$ and the corresponding hole has a position $ \geq k$, resulting in an unpaired minus sign. As a consequence, ${\cal O}_P^2(l)$ decreases with increasing $J/U$ since the probability to cut particle-hole pairs becomes larger. Finally, at the transition to the SF phase, the pairs begin to deconfine and overlap, resulting in a completely random product of signs and $ {\cal O}_P^2=0$.

\begin{figure}[!t]
\begin{center}
\includegraphics[width=0.9\columnwidth]{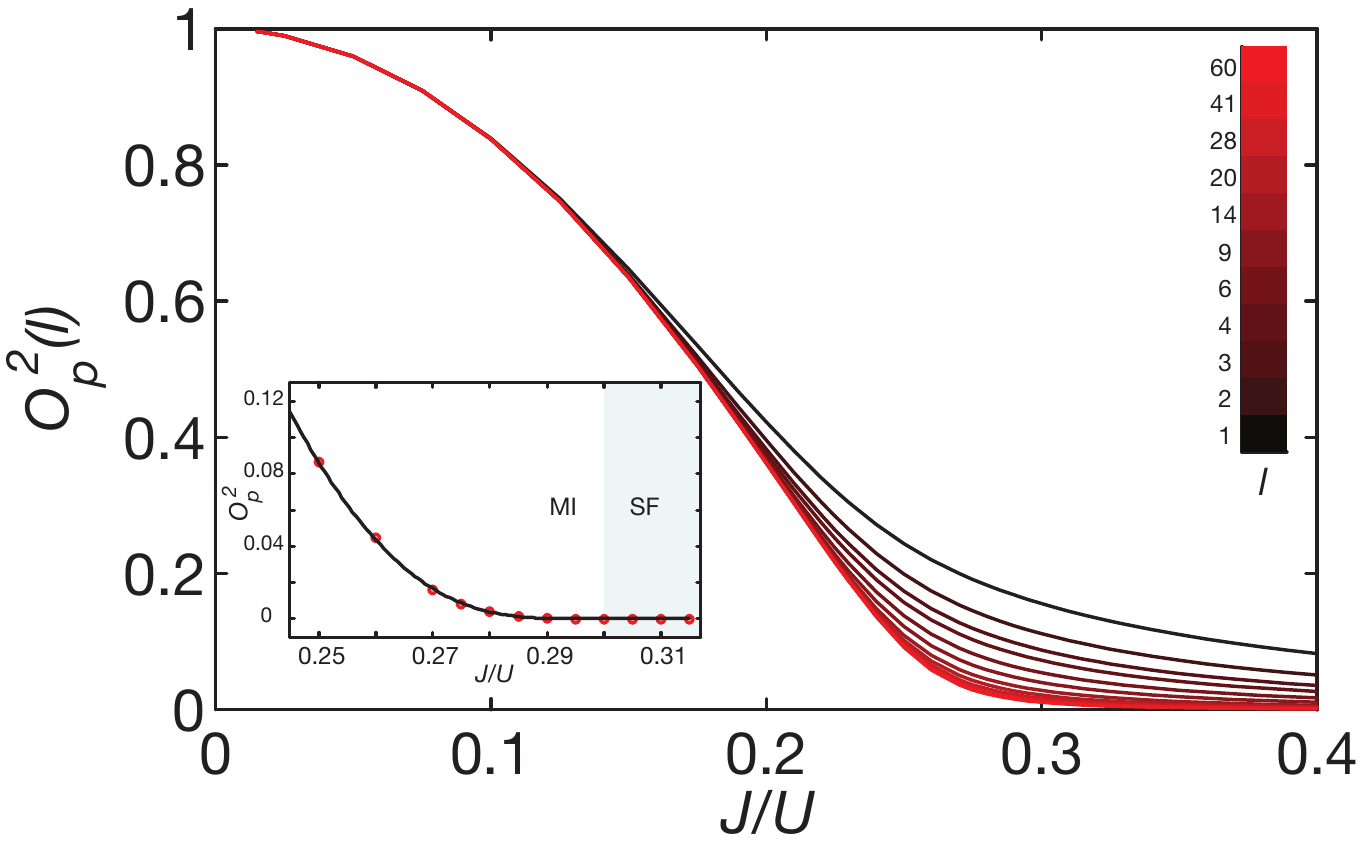}
\end{center}
\vspace{-0.4cm}
\caption{ {\bf Numerical calculation of the string-order parameter.}  ${\cal O}_P^2(l)$ as a function of $J/U$ calculated with DMRG for a homogeneous chain  ($\bar n = 1$, $T = 0$) of total length $216$. Lines show ${\cal O}_P^2(l)$ for selected lengths $l$ (black to red colors). Inset: Extrapolated value ${\cal O}_P^2=\lim_{l\rightarrow\infty}{\cal O}_P^2(l)$ together with a fit (black line) of the form  \mbox{$ {\cal O}_P^2\propto \exp \big(-A\left [(J/U)_c^{1d}-(J/U)\right ]^{-1/2}\big)$}, characteristic for a transition of BKT type (see Appendix). \label{fig:theory}}
\end{figure}

To support this intuitive argument, we calculated ${\cal O}_P^2(l)$ numerically using DMRG in a homogeneous system at $T=0$. We show ${\cal O}_P^2(l)$ in Fig.\,\ref{fig:theory} for selected distances  $l$ together with the extrapolated values of  ${\cal O}_P^2=\lim_{l \rightarrow \infty} {\cal O}_P^2(l)$ (inset to Fig 3), which we computed using finite-size scaling  (see Appendix). We performed a fit to the extrapolated values close to the critical point with an exponential scaling 
\begin{equation}\label{eq:BKTScaling}
{\cal O}_P^2\propto \exp \big(-A{\left [ (J/U)_c^{1d}-(J/U)\right]^{-1/2}}\big)
\end{equation}
characteristic for a transition of BKT type (Fig.\,\ref{fig:theory})  that one expects in one dimension \cite{Berg:2008, Kuehner:1998}. From the fit we find  $(J/U)_c^{1d}=0.295-0.320$, which is compatible with previously computed values  \cite{Kuehner:2000,Kashurnikov:1996a} (see Appendix). The fact that $ {\cal O}_P=0$ in the SF and $ {\cal O}_P>0$ in the MI as well as the agreement with the expected scaling show that $ {\cal O}_P$ serves as an order parameter for the MI phase in one dimension \cite{Berg:2008}. Additionally, the simulations demonstrate that ${\cal O}_P(l)$ is well suited to characterize the SF-MI transition even for finite lengths $l$.


\begin{figure}[b]
\begin{center}
\includegraphics[width=\columnwidth]{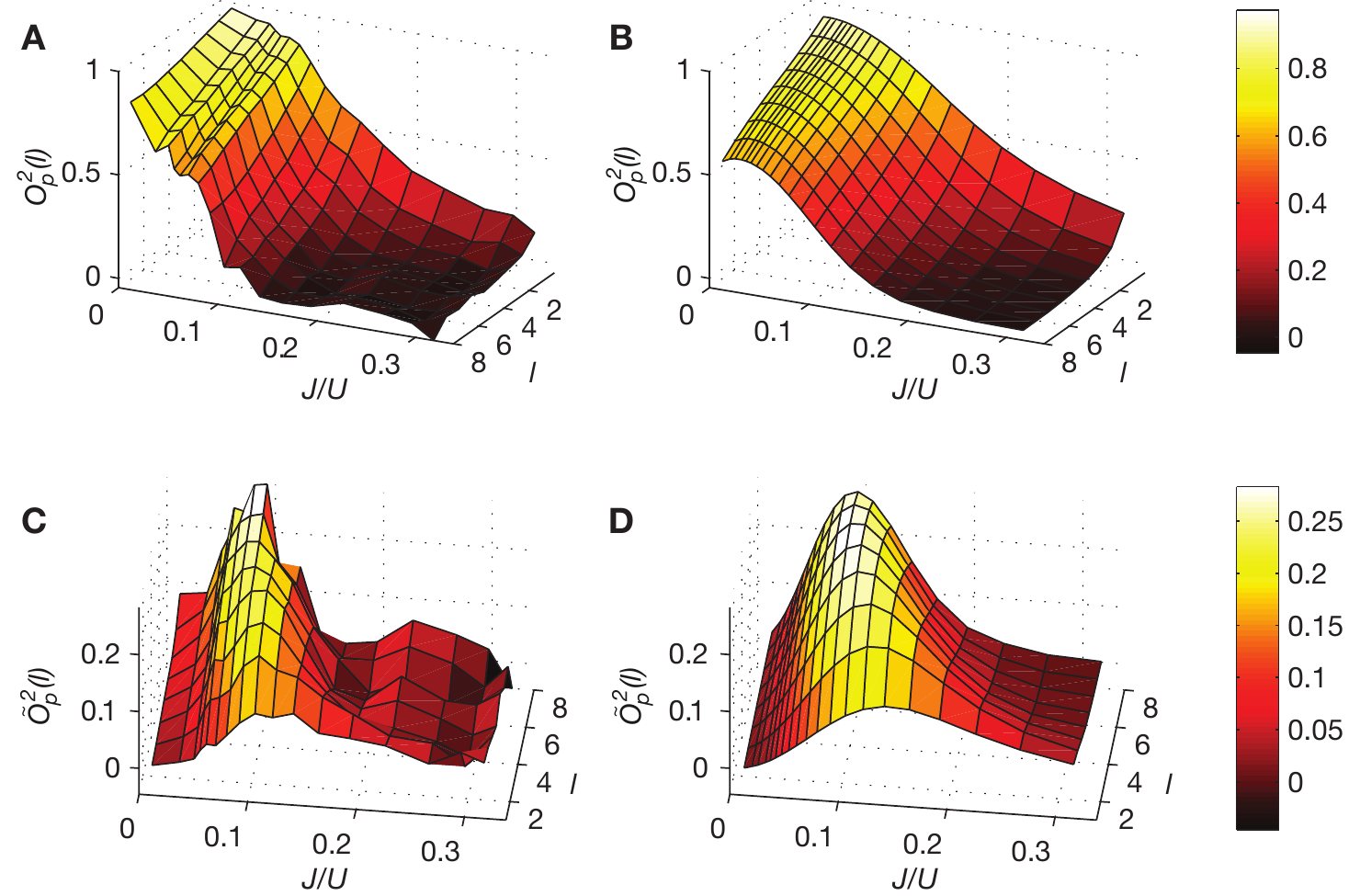}
\end{center}
\vspace{-0.5cm}
\caption{ {\bf String correlators.} ({\bf A}) Experimental values of ${\cal O}_P^2(l)$ for lengths $0 \leq l \leq 8$  ({\bf B}) In-trap MPS calculations at $T=0.09\,U/k_B$. ({\bf C}) Experimentally determined string correlator $\tilde {\cal O}_P^2(l)$ as defined in Eq.\,\ref{eq:Opl_tilde} for lengths $1 \leq l \leq 8$ ({\bf D}) In-trap MPS calculations at $T=0.09\,U/k_B$. The $l$ axes in (C) and  (D) have been inverted.
\label{fig:string}}
\end{figure}

Our experimentally obtained values of  ${\cal O}_P^2(l)$ for string length $l \leq 8$ (Fig.\,\ref{fig:string}A) agree qualitatively well with in-trap MPS calculations at $T=0.09\,U/k_B$ (Fig.\,\ref{fig:string}B).  We observe a stronger decay of ${\cal O}_P^2(l)$ with $l$ compared to the $T=0$ case, because at finite temperature thermal fluctuations lead to minus signs at random positions of the chain and reduce the average value of ${\cal O}_P^2(l)$. Despite that, we still see a strong growth of ${\cal O}_P^2(l)$, once the transition from the SF to the MI is crossed, with a similar behavior as in Fig.\,\ref{fig:theory}.


For a completely uncorrelated state, ${\cal O}_P^2(l)$ factorizes to $\prod_{k\leq j \leq k+l}\avs{\hat s_j}$, and in a homogeneous system we would expect a decay with string length of the form $\avs{\hat s_j}^{l+1}$, which can be slow provided the mean on-site parity $\avs{\hat s_j}$ is close to one. To rule out that our experimental data shows only such a trivial behavior, we define a new quantity $\tilde{\cal O}_P^2(l)$ that more naturally reflects the underlying correlations:
\begin{equation}
\tilde{\cal O}_P^2(l)={\cal O}_P^2(l)- \prod_{k\leq j \leq k+l}\avs{\hat s_j}.\label{eq:Opl_tilde}
\end{equation}
First, we notice that $\tilde{\cal O}_P^2(l)$ for length $l=1$ is equal to the two-site correlation function $C(d=1)$. Second,  $\tilde{\cal O}_P^2(l)\approx{\cal O}_P^2(l)$  for long distances $l$ since $\prod_{k\leq j \leq k+l}\avs{\hat s_j}$ eventually decays to zero (except for the singular case $J/U=0$ and $T=0$). The correlation function $\tilde{\cal O}_P^2(l)$ can therefore be understood as an extension of the two-site correlation function, that essentially captures the physics behind string order in one-dimensional MIs.

Experimental and theoretical values for $\tilde{\cal O}_P^2(l)$ are shown in Figs.\,\ref{fig:string}C and D. For small $J/U$, $\tilde{\cal O}_P^2(l)$ is significantly reduced compared to ${\cal O}_P^2(l)$ since few particle-hole pairs exist and ${\cal O}_P^2(l)$ is close to its factorized form for short lengths $l$. In the case of vanishing $J/U$, we even expect $\tilde{\cal O}_P^2(l)=0$ since all sites are completely decoupled. For intermediate $J/U\approx0.1$,  $\tilde{\cal O}_P^2(l)$ grows rapidly with length $l$ showing a strong deviation from the factorized form. Finally, in the SF regime, $ \tilde{\cal O}_P^2(l)$ becomes indiscernible from zero for large lengths in contrast to the nearest-neighbor two-site correlation function. Furthermore, our data shows a genuine three-site correlation, which we revealed after subtracting all two site correlators in addition to local terms (see Appendix).

We have shown direct measurements of non-local parity-parity correlation functions on the single-lattice-site and single-atom level and we demonstrated that a one-dimensional Mott insulator is characterized by non-local string order. A natural extension of our work would be to reveal, e.g., topological quantum phases such as the Haldane insulator of bosonic atoms \cite{DallaTorre:2006,Berg:2008}. A Haldane insulator exhibits a hidden antiferromagnetic ordering and is expected to occur in one-dimensional quantum gases in the presence of longer ranged interactions, which could be realized in our experiment using Rydberg atoms \cite{Saffman:2010}.


\section*{Acknowledgements}
 We acknowledge helpful discussions with Ehud Altman, Emanuele Dalla Torre, Matteo Rizzi and Ignacio Cirac. This work was supported by MPG, DFG, EU (NAMEQUAM, AQUTE, Marie Curie Fellowship to M.C.), and JSPS (Postdoctoral Fellowship for Research Abroad to T.F.).  L.P. is supported by the Swiss National Science Foundation under grant PZ00P2-131892/1.

\bibliography{ParticleHolePreprint}


\section*{Appendix}
\renewcommand{\arraystretch}{1.5}

\subsection{Preparation of a two-dimensional degenerate quantum gas}
Most of the experimental details are the same as in Refs.\,\cite{Sherson:2010} and \cite{Weitenberg:2011}, but several improvements have been implemented in order to increase the stability of the system. We started by transporting a thermal cloud of $^{87}$Rb atoms in the hyperfine state $\ket{F=1, m_F=-1}$ with a single-beam optical dipole trap (wavelength 1064\,nm, beam waist $w_0=40\,\mu$m) in front of the high-resolution imaging system. Afterwards, the atoms were loaded into a crossed dipole trap, which consists of the horizontal lattice beams (wavelength 1064\,nm, beam waist $w_0=70\,\mu$m) whose retro-reflections were blocked with mechanical shutters. We evaporatively cooled the atoms in this trap by changing its depth from $20\,\mu$K to $10\,\mu$K within $1$\,s, before transferring the atoms into a vertical standing wave (wavelength 1064\,nm, beam waist $w_0=70\,\mu$m) of which we typically populate $40$ antinodes.

To extract a single two-dimensional system, we used a position-dependent microwave transfer in a magnetic field gradient of $\partial B/\partial z = 45$\,G/cm, 1.9 times larger compared to Ref.\,\cite{Sherson:2010}. After a transfer of all atoms from $\ket{F=1, m_F=-1}$ to $\ket{F=2, m_F=-2}$ with a broad microwave sweep, the atoms from one antinode of the standing wave were transferred back to $\ket{F=1, m_F=-1}$ using a HS1-pulse \cite{Weitenberg:2011} of $5$\,ms duration and a sweep width of $3.5$\,kHz. The remaining atoms in state $F=2$ were removed from the trap by a laser pulse resonant with the $F=2\rightarrow F'=3$ transition.

For the final evaporation, we added a focussed laser beam of wavelength 850\,nm shone in through the imaging system, with a beam diameter of $\sim\!20\,\mu$m at the position of the atoms. The intensity of the beam is adjusted in such a way that a small number of atoms is lost due to three-body recombination, resulting in a stabilization of the atom number. In this configuration, we evaporatively cooled the atoms by tilting the trap horizontally with a magnetic field gradient of $11.25$\,G/cm and by reducing the power of the 850\,nm beam. We adjusted the atom number by changing the end point of this evaporation and achieved an atom number stability better than $20$\,\%.

\subsection{Experimental sequence and lattice calibration}
For the measurements on the one-dimensional systems, we changed the potential depth of lattice axis $y$ to a fixed final value $V_y = 17(1)\,E_r$ using an s-shaped ramp of 120\,ms duration after having created a degenerate quantum gas in a single antinode of the vertical lattice beam. At the same time and using the same ramp shape, we ramped the lattice depth along $x$ to variable final depths $V_x$ to attain different values of  $J/U$. The vertical lattice depth was kept at $V_z = 20(1)\,E_r$ during these ramps. For the measurement of two-site correlations in two dimensions, the depths of lattice axis $x$ and $y$ were changed simultaneously to the same final value.

We obtained $J$ and $U$ from the lattice depths by a numerical band-structure calculation. In order to calibrate the lattice depths, we modulated the lattice intensity and measured parametric excitations of the atoms from the zeroth to the second Bloch band. As a result, we observed a resonance in the in-situ width of the atom cloud as a function of the modulation frequency. Our method allowed us to measure the transition frequency with an uncertainty of $2\%$. Besides this, we saw drifts of the lattice depth during our measurements of typically $5\%$. Both errors together lead to an experimental uncertainty of $\sim\,10\%$ for $J/U$.

For the detection, we froze the density distribution of the atoms by increasing all axes simultaneously
to $V_x=72\,E_r$, $V_y=78\,E_r$, and $V_z=83\,E_r$ within $0.2$\,ms.  We checked that our ramps were fast enough  to avoid any local number-squeezing dynamics \cite{Bakr:2010}. Finally, the individual atoms on the lattice were detected using a high-resolution objective with numerical aperture $\mbox{NA}=0.68$. For this, we further increased the lattice depths along all three directions to $\sim\,3000\,E_r$ within $2$\,ms and illuminated the atoms with an optical molasses that induces fluorescence and simultaneously laser cools the atoms \cite{Sherson:2010}.

\subsection{Bose-Hubbard model and perturbation theory}
For the temperatures and lattice parameters relevant to the presented measurements, the physics of a gas of interacting bosons in a lattice is captured by the Bose-Hubbard model \cite{Jaksch:1998}:
\begin{equation}
\hat{H} = -J \sum_{<i,j>} \hat{a}^{\dagger}_i \hat{a}_{j} + \frac U2 \sum_i \hat{n}_i(\hat{n}_i-1)+\sum_i \epsilon_i \hat{n}_i
\end{equation}
Here, $\hat{a}_i$ and $\hat{a}_i^\dagger$ correspond to the bosonic annihilation and creation operators at site $i$, $\hat{n}_i=\hat{a}_i^\dagger\hat{a}_i$ is the on-site number operator, and $\epsilon_i$ denotes the energy offset due to an external harmonic confinement.

The emergence of quantum-correlated particle-hole pairs in a Mott insulator with finite tunnelling can be readily understood within first-order perturbation theory. For a Mott insulator in the atomic limit, where the ratio of the tunneling energy $J$ to the on-site interaction energy $U$ vanishes, the many-body state is simply given by $|\Psi\rangle_{J/U=0}=\prod_i |n_i=1\rangle$ for the case of unity filling and $T=0$. For finite but still small tunnelling, $J/U \ll 1$, the ground state can be approximated by considering the tunnelling term as a perturbation. To first order, one obtains:
\begin{equation}\label{eq:PertTheory}
  |\Psi\rangle_{J/U \ll 1}
\propto
   |\Psi\rangle_{J/U=0} + \frac{J}{U}\sum_{\langle i,j \rangle} \hat a_i^\dagger \hat a_j |\Psi\rangle_{J/U=0},
\end{equation}
which yields a probability to find a particle-hole pair on neighboring sites proportional to $(J/U)^2$. Within the approximation of Eq.\,\ref{eq:PertTheory} the nearest-neighbor parity-correlation function is given by $C(d=1)=16 (J/U)^2+\mathcal{O}[(J/U)^4]$. Closer to the transition to the superfluid and for large system sizes, higher-order perturbation terms become more  important and essentially lead to a rapid increase of bound particle-hole pairs and an extension of their size, eventually resulting in deconfinement of the pairs at the transition point.

\subsection{On-site variance and next-nearest-neighbor correlations}
\begin{figure}[!t]
\begin{center}
\includegraphics[width=\columnwidth]{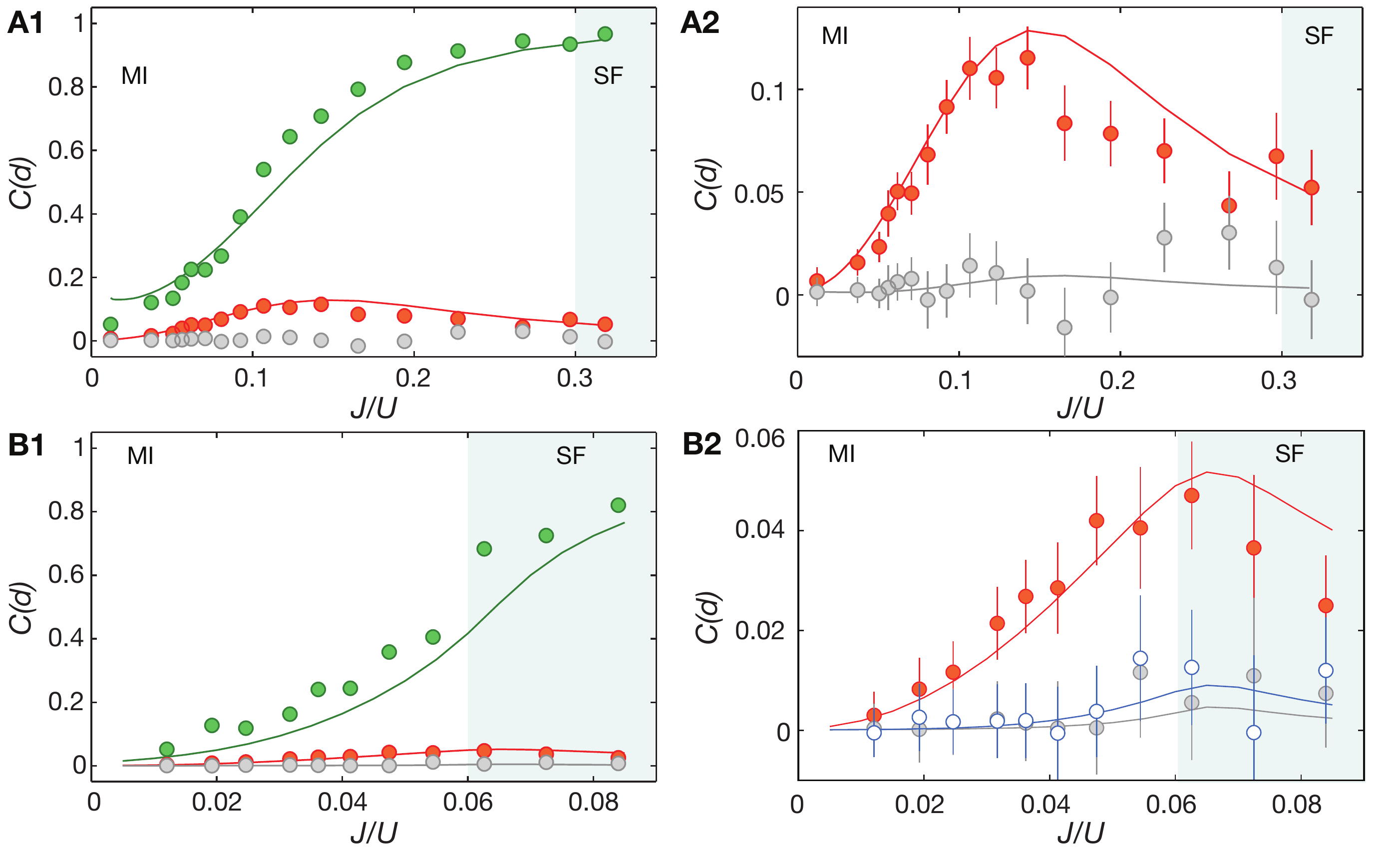}
\end{center}
\vspace{-0.5cm}
\caption{ {\bf On-site variance and next-nearest-neighbor correlations.} ({\bf A1})
 Parity correlations $C(d)$ for different distances $d=0,1,2$ (green, red and gray circles, respectively) for the one-dimensional systems. The solid lines are finite-temperature MPS calculations including harmonic confinement at $T=0.09\,U/k_B$. ({\bf A2}) Enlarged view of $C(d)$ for $d=1,2$ for the same datasets as (A1). ({\bf B1}) Same quantities as in (A1),  but for the two-dimensional case whereas the data is the mean of $C(d)$ for both possible directions. ({\bf B2}) Enlarged view of $C(d)$ for $d=1,2$ (red, gray) and additionally the two-site correlations for the next-nearest-neighbor along the diagonal (blue). The solid lines are a QMC calculation for a homogeneous system at $T=0.1\,U/k_B$. The statistical errorbars in (A1) and (B1) are smaller than the dot size. We attribute the systematic shift of $C(d=0)$ in (B) to the inhomogeneous trapping potential.
\label{fig:pairDIST}}
\end{figure}

In addition to the nearest-neighbor parity correlation $C(d=1)$, we also evaluated the correlations
$C(d)$ (Eq.\,\ref{eq:Cd}) for distances $d=0$ and $d=2$ in our one-dimensional systems (Fig.\,\ref{fig:pairDIST}-A1). For $d=0$, this amounts to measuring the on-site variance $C(d=0) = \sigma(\hat s_k)^2$. Deep in the Mott-insulating regime, the on-site number distribution is strongly squeezed and the variance is close to zero, whereas in the SF regime, it saturates at $\sigma(\hat s_k)^2=1$, as expected for the case of a parity detection\footnote{In Ref.\,\cite{Sherson:2010} we used a different definition of the parity operator, $\hat p_k = (\hat s_k +1)/2$, which yielded 1(0) for an odd (even) occupation number. In that case, the maximum variance was
$\sigma(\hat p_k)^2=\sigma(\hat s_k)^2/4 = 0.25$.}. $C(d)$ drops rapidly as a function of the distance $d$ and the numerical calculations predict only a small maximum of $0.01$ in the next-nearest-neighbor correlation $C(d=2)$ at $J/U\sim0.17$, which is however indiscernible from the statistical noise in our measurements (Fig.\,\ref{fig:pairDIST}-A2).

We also show the parity correlation $C(d)$ in the two-dimensional system  for different distances
(Fig.\,\ref{fig:pairDIST}-B1). We found that, similar to the one-dimensional systems, next-nearest-neighbor correlations are not visible above the statistical noise, whereas the QMC simulations show a small maximum around $J/U\sim0.065$ (Fig.\,\ref{fig:pairDIST}-B2). However, a striking difference is that the on-site fluctuations around the critical point are significantly smaller than in one dimension.

\subsection{String order and multi-site correlations}
In the following paragraph, we show that our data for the string-type correlators (Fig.\,\ref{fig:string}) cannot be explained with pure two-site correlations. Additionally, we address the connection between string order and multi-site correlations.

To illustrate this, let us consider the simplest case of a string-type correlator including three sites $\avs{\hat{s}_1\hat{s}_2\hat{s}_3}$, where $\hat{s}_1$,$\hat{s}_2$ and $\hat{s}_3$ refer to the parity of three neighboring sites on a one-dimensional chain. If one of the sites is not correlated with the others, we can calculate $\avs{\hat{s}_1\hat{s}_2\hat{s}_3}$ from two-site and on-site terms. For instance, if site $3$ is not correlated with sites $1$ and $2$, we have $\avs{\hat{s}_1\hat{s}_2\hat{s}_3}=\avs{\hat{s}_1\hat{s}_2}\avs{\hat{s}_3}$. This fact can be generally  expressed  using a three-site cumulant $\avs{\hat{s}_1\hat{s}_2\hat{s}_3}_c$, defined as \cite{Kubo:1962}:
\begin{eqnarray}
 \avs{\hat{s}_1\hat{s}_2\hat{s}_3}_c&=&\avs{\hat{s}_1\hat{s}_2\hat{s}_3}-\avs {\hat{s}_1}\avs{\hat{s}_2}\avs{\hat{s}_3}\nonumber\\
 &&-C_{1,2}\avs{\hat{s}_3}-C_{2,3}\avs{\hat{s}_1} -C_{1,3}\avs{\hat{s}_2} \label{eq:fullthree}
\end{eqnarray}
with two-site correlation functions $C_{i,j}=\avs{\hat{s}_i\hat{s}_j}-\avs{\hat{s}_i}\avs{\hat{s}_j}$. The cumulant is a measure of the correlations between all three sites as it vanishes if one of the sites is not correlated with the others. Additionally, if the cumulant vanishes,  Eq.\,\ref{eq:fullthree} can be written as $\avs{\hat{s}_1\hat{s}_2\hat{s}_3}=\avs {\hat{s}_1}\avs{\hat{s}_2}\avs{\hat{s}_3}+C_{1,2}\avs{\hat{s}_3}+C_{2,3}\avs{\hat{s}_1} +C_{1,3}\avs{\hat{s}_2}$ and we therefore have a situation where $\avs{\hat{s}_1\hat{s}_2\hat{s}_3}$ does not contain more information than two-site and on-site terms.

\begin{figure}[b]
\begin{center}
\includegraphics[width=0.75\columnwidth]{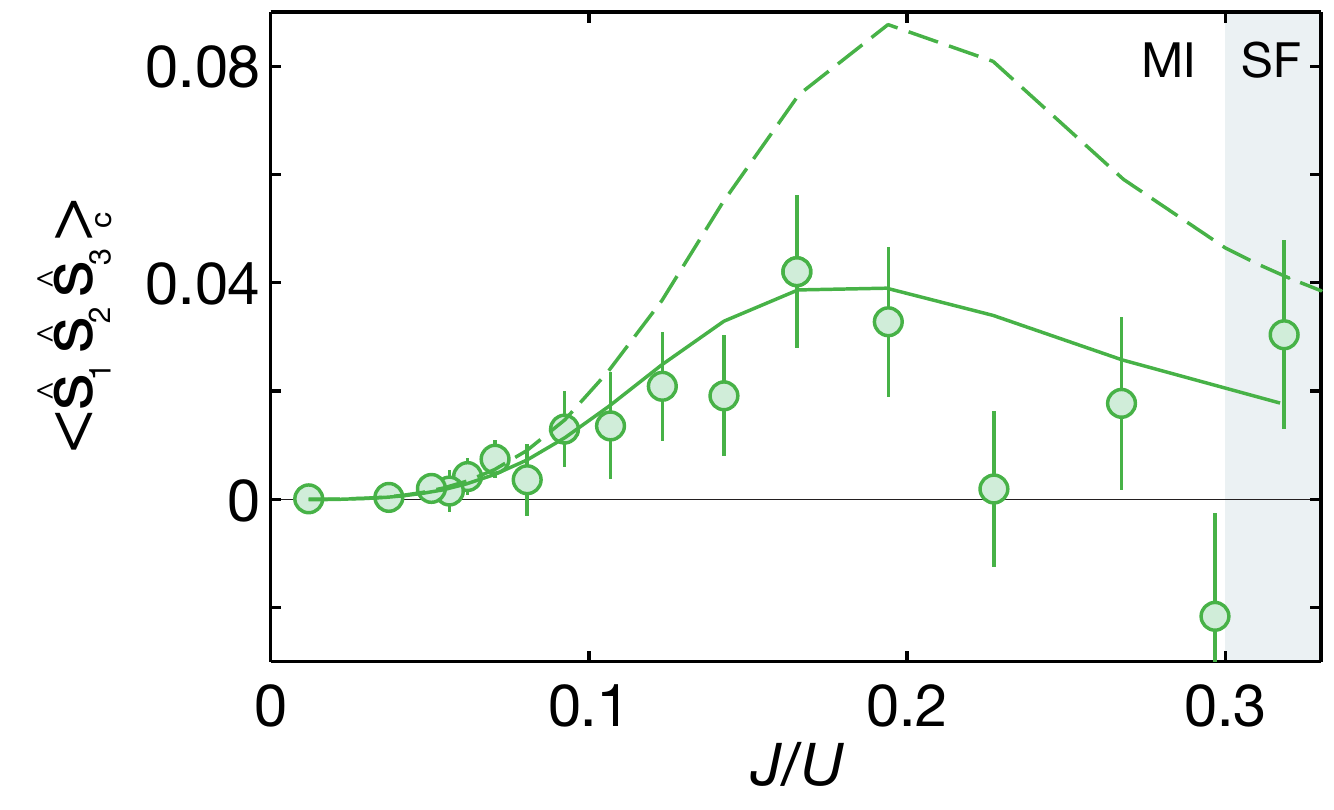}
\end{center}
\vspace{-0.5cm}
\caption{{\bf Three-site correlator.} Our experimental values for three site cumulant $\avs{\hat{s}_1\hat{s}_2\hat{s}_3}_c$ (green circles) show the existence of three-site correlations in our system.   The curves are DMRG calculations for a homogeneous system at $T=0$ (dashed line) and finite-temperature MPS calculations including harmonic confinement at $T=0.09\,U/k_B$ (solid line).
\label{fig:fullthree}}
\end{figure}

Our experimental values for the three-site cumulant $\avs{\hat{s}_1\hat{s}_2\hat{s}_3}_c$ (Fig.\ref{fig:fullthree}) show a significant signal for $J/U<0.20$, in quantitative agreement with a MPS calculation at $T=0.09\,U/k_B$ including the harmonic confinement. Note that this is a substantial effect, as the peak value constitutes about $40\%$ of the peak value of $\tilde{\cal O}_P^2(l=2)$, where only on-site terms have been subtracted. We draw two conclusions from this: First, our data shows a three-site correlation beyond simple two-site correlations. Second, our data for string-type expectation values cannot be expressed in terms of two-site correlations alone.
Let us explain the latter statement in more detail. It turns out that if a three-site expectation value cannot be expressed via two-site terms, then expectation values including more than three sites cannot be expressed in this way either. The four-site term, e.g., can be written as:
\begin{eqnarray}
\avs{\hat{s}_1\hat{s}_2\hat{s}_3\hat{s}_4}  & =& \avs{\hat{s}_1\hat{s}_2\hat{s}_3\hat{s}_4}_c+\avs{\hat{s}_1\hat{s}_2\hat{s}_3}_c\avs{\hat{s}_4}+ \avs{\hat{s}_2\hat{s}_3\hat{s}_4}_c\avs{\hat{s}_1}\nonumber \\
&&+ \avs{\hat{s}_1\hat{s}_2\hat{s}_4}_c\avs{\hat{s}_3}+\avs{\hat{s}_1\hat{s}_3\hat{s}_4}_c\avs{\hat{s}_2}+C^{(2)} \label{eq:fullfour}
\end{eqnarray}
where $C^{(2)}$ contains only terms including two-site and on-site terms. Even if there were no correlations between four sites $\avs{\hat{s}_1\hat{s}_2\hat{s}_3\hat{s}_4}_c=0$, we still had to include the non-vanishing  three-site cumulants. Therefore $\avs{\hat{s}_1\hat{s}_2\hat{s}_3\hat{s}_4}$ cannot be expressed via two-site and on-site terms. The argument can easily be extended to longer strings and we conclude that our signal for string-type correlators including more than two-sites cannot be explained with pure two-site correlations.

In general, every string-type correlator can be expressed as a sum of products of cumulants similar to Eq.\,\ref{eq:fullfour}. Since cumulants are a measure of multi-site correlations, such an expansion gives us information about the contribution of multi-site correlations to string order. In the atomic limit where $J/U\approx0$, even two-site correlations are absent, but string order is present.  In this case, the string signal is trivially dominated by on-site terms. For small but finite $J/U$ values, three-site correlations almost vanish and the string signal is dominated by two-site correlations and on-site terms. In the range of $J/U\approx0.1-0.2$, three-site correlations build up and also contribute to the string-order signal. Generally, it would be  interesting to know how this relation extends when approaching the critical point. One particularly interesting question is whether close to the critical point multi-site correlations between an infinite number of sites exist in the thermodynamic limit. Such an analysis is difficult  theoretically and experimentally, which can already be seen from the expression for the four-site cumulant. This is however beyond the scope of this manuscript and a matter of further investigation.

Instead, we continue with an interpretation of the present three-site correlations. One explanation for such a signal would be the existence of pairs extending over three sites. However, another possibility can explain such correlations: Assume we restrict ourselves to a system of only three sites, where the parity on each site can take the values $s_i=\pm1$, $i=1,2,3$.  The system is fully characterized by the probability $p(s_1,s_2,s_3)$ of finding the parities $s_1,s_2$ and $s_3$ on sites $1$,$2$ and $3$. In a situation, where only nearest-neighbor pairs exist with probability $p_{nn}$, we have $p(+,+,+)=1-2p_{nn}$, $p(-,-,+)=p(+,-,-) =p_{nn}$ and $p(s_1,s_2,s_3)=0$ for all other cases. Interestingly, we find a non-vanishing three-site cumulant  $\avs{\hat{s}_1\hat{s}_2 \hat{s}_3}_c=16 p_{nn}^2-32 p_{nn}^3$ in this situation. The three-site correlation remains present even if we consider the three sites as a subsystem of a longer chain. We conclude that a three-site correlation can arise from next-neighbor pairs alone, simply because site $1$ is correlated with site $2$ and site $2$ is correlated with site $3$. However, our signal extends far beyond the region where first order perturbation theory is valid and we can therefore assume that particle-hole pairs with an extension of more than two sites as well as more complicated clusters play a significant role in our situation.

Finally, we would like to point out an important difference between two-site correlators and string correlators: Averaging over many experimental realizations of the system, a two-site correlator at distance $d$ sums up correlations from pairs with exactly an extension of $d$ but also anti-correlations from all pairs with a different size.  In contrast, a string-type correlator of length $l$ sums up positive contributions from all pairs which have a size less than $l$ and lie within the string length. This feature makes the string-type correlator more suitable to study the Mott insulating phase. In particular, the phase transition in one dimension is marked by the vanishing of the string correlators for long lengths. It is not possible to extract the same information using only two-site correlators.

\subsection{Comparison of string correlator with theory}
\begin{figure}[!h]
\begin{center}
\includegraphics[width=0.8\columnwidth]{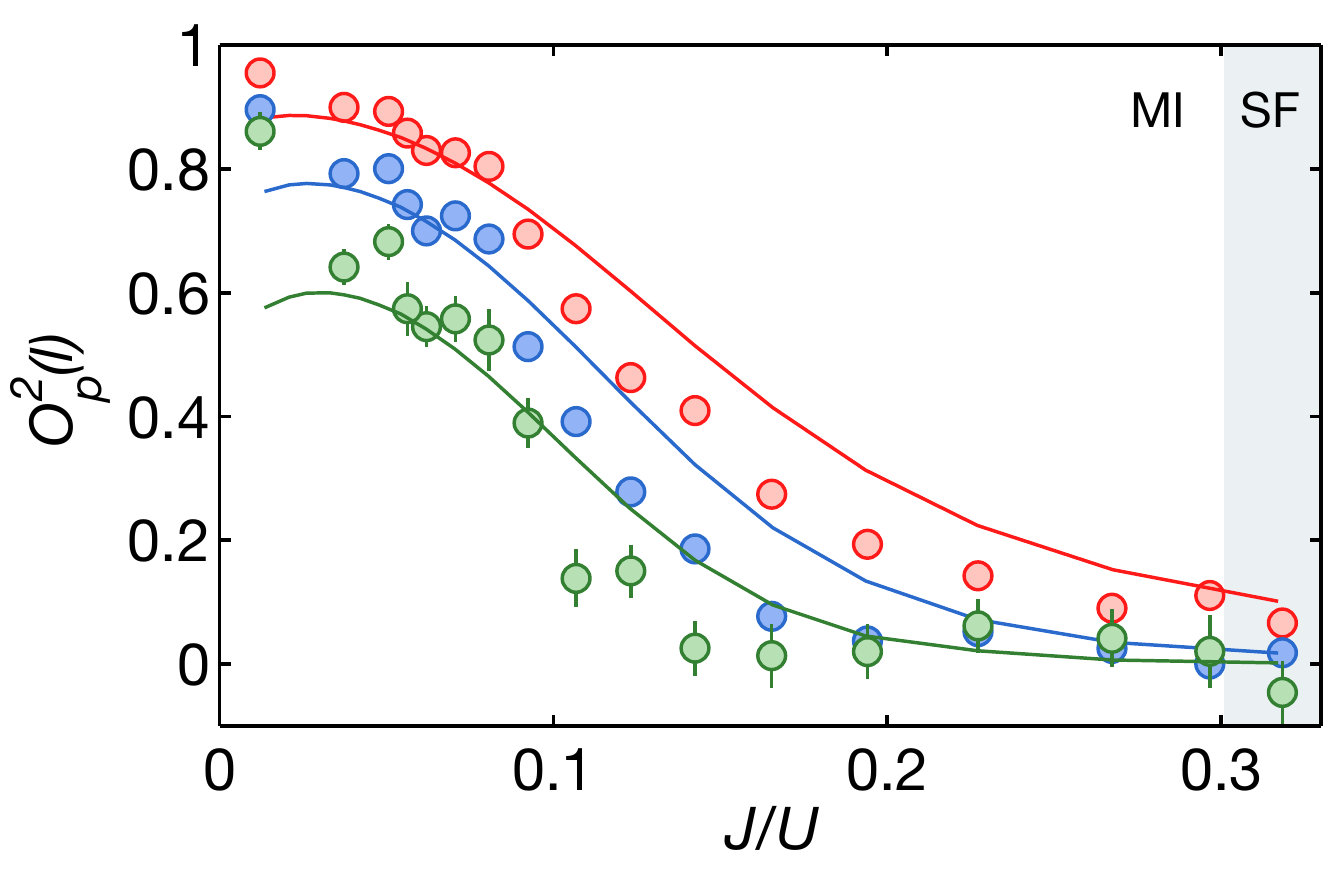}
\end{center}
\vspace{-0.5cm}
\caption{  {\bf String correlator - comparison with theory.}
String correlator ${\cal O}_P^2(l)$ for different lengths $l=1$ (red circles), $l=4$ (blue circles) and $l=8$ (green circles).
These curves are cuts of the three-dimensional representation of the data, as shown in Fig.\,\ref{fig:string}.
Solid lines correspond to  finite-temperature MPS calculations including harmonic confinement at $T=0.09\,U/k_B$ with the same color coding as the experimental data.
\label{fig:cuts}}
\end{figure}

For a more detailed comparison of the experimentally obtained  string correlations with theory (Fig.\,\ref{fig:string}), we show ${\cal O}_P^2(l)$ for different lengths $l=1$ (Fig.\,\ref{fig:cuts}, red circles), $l=4$ (blue circles) and $l=8$ (green circles) together with MPS calculations at $T=0.09\,U/k_B$ including the harmonic confinement. We observe a good qualitative agreement of the data with the numerical simulations. Systematic errors of the experimental data can arise from different mean atom numbers for different $J/U$, leading to different local chemical potentials. A systematic discrepancy between theory and experiment can result from a small mismatch of the trapping frequencies or the calibration of $J$ and $U$. Additionally, the theoretical prediction is calculated at fixed temperature (in units of $U$), but the experimental data is approximately taken at a constant entropy. In the latter case, we expect the temperature to scale with $U$ for small $J/U$ values, but it is not clear that this scaling remains valid for higher $J/U$ values. The temperature for the numerical simulation was chosen to yield the best agreement over the full range of $J/U$ values.


\subsection{DMRG and finite-size scaling}
We use a DMRG code with open boundary conditions, which implements the conservation of the number of particles. The considered system sizes span between $L=168$ and $L=256$. The number of retained states $m$ has been chosen in order to have a truncation error smaller than $10^{-8}$\cite{schollwoeck:2005}. This results in a relative error for the shown data which is smaller than $10^{-3}$, as estimated via comparison of data for different $m$ between $185$ and $235$.

The string correlator is strongly affected by the presence of boundaries and finit- size effects. We introduce the notation $O_P^2 (J/U,l,L)$ to address the expectation value of the string correlator of length $l$ obtained from the numerical simulation of a system of length $L$. The $l$ sites are always taken in the center of the system to minimize boundary effects.
For the finite size scaling in the inset of Fig.\,\ref{fig:theory}, we analyzed correlators $O_P^2 (J/U, \alpha L, L)$ for a fixed fraction $\alpha$ of the total length $L$ ($l=\alpha L$). We extrapolated $O_P^2 (J/U, \alpha) = \lim_{L \rightarrow \infty} O_P^2 (J/U,\alpha L,L)$ using a scaling of the form $O_P^2 (J/U,\alpha L,L)=a+b/L^\eta$ \cite{Kennedy:1992, Ueda:2008, Dalmonte:2011}.
\begin{table}[!h]
\begin{center}
\begin{tabular}{|c||c|c|c|}
\cline{1-4} & $\alpha=1/4$   & $\alpha=1/3$ & $\alpha=1/2$ \\
\hhline{|=||=|=|=|} $[(\frac{J}{U})_1,(\frac{J}{U})_2]=[0.23,0.37]$ & $0.303$ &  $0.306$ & $0.296$ \\
 \hline$[(\frac{J}{U})_1,(\frac{J}{U})_2]=[0.24,0.36]$ &  $0.310$&  $0.311$ &  $0.299$ \\
\hline $[(\frac{J}{U})_1,(\frac{J}{U})_2]=[0.25,0.35]$ &  $0.319$&  $0.318$ &  $0.305$ \\
\hline $[(\frac{J}{U})_1,(\frac{J}{U})_2]=[0.26,0.34]$ &  $0.321$ &  $0.319$ &  $0.311$ \\
\hline  $[(\frac{J}{U})_1,(\frac{J}{U})_2]=[0.27,0.33]$ &  $0.317$ &  $0.314$ &  $0.309$ \\
\hline
\end{tabular}
\end{center}
\vspace{-0.4cm}
\caption{Results for $(J/U)_c$ for different relative lengths $\alpha$ and fitting intervals $[(\frac{J}{U})_1,(\frac{J}{U})_2]$. The  fitting error for a given $\alpha$ and $[(\frac{J}{U})_1,(\frac{J}{U})_2]$ is below $0.003$.}
\label{tab:juc}
\end{table}

To determine $(J/U)_c$, we fitted the extrapolated values with  ${\cal O}_P^2\propto\exp \big(-A{\left [ (J/U)_c^{1d}-(J/U)\right]^{-1/2}}\big)$. The results for  $(J/U)_c$ appear to be strongly dependent on the fitting interval $[(J/U)_1,(J/U)_2]$ and $\alpha$, as shown in Table \ref{tab:juc}. This large systematic error could be reduced using more advanced finite size-scaling methods \cite{Kashurnikov:1996b,Ueda:2008}, which is beyond the scope of this manuscript. In the inset of Fig.\,\ref{fig:theory}, we show the fit using $\alpha=1/2$ and $[(J/U)_1,(J/U)_2]=[0.25,0.35]$ .

\subsection{Finite-temperature MPS}
Thermal equilibrium states for a finite chain of $q$-dimensional
systems can be approximated by MPS using the techniques introduced
in~\cite{Verstraete:2004,Zwolak:2004}.  For a chain of length $L$, the density
matrix can be expressed as a vector in a Hilbert space of larger
dimension, $q^{2L}$. We approximate this by a MPS of length $2 L$,
where sites $2k$ and $2k+1$ correspond to the $k$-th site in the
physical system.
For inverse temperature $\beta=1/(k_B T)$, the thermal state (up to
normalization) is formally identical to the imaginary time evolution
of the identity operator $\rho\approx e^{-\beta H}=e^{-\frac{\beta}{2}
H} \mathbb 1 e^{-\frac{\beta}{2} H}$.
To obtain the MPS representation of this operator,
we use a second-order Suzuki-Trotter decomposition of the exponentials,
and apply the imaginary time evolution to the initial state
corresponding to the identity operator \cite{Verstraete:2004}.

Using this method, we have simulated the thermal states of a Bose-Hubbard
chain with length $L=30$, for up to $n=4$ particles per site. We used an average over systems with different total atom numbers $N$, corresponding to the typical atom number distribution in the experiment, as well as an average over the central $9$ sites. The error of the numerical simulation originates from the Trotter step $\delta$
and the bond dimension parameter $D$, which controls the
number of parameters in the MPS ansatz. Both errors can be reduced
by running simulations with increasing (decreasing) $D$ ($\delta$).
We have used $D=20$, $\delta=0.05$, for which the estimated error is below
$10^{-3}$.

\subsection{Quantum Monte Carlo calculations}
Numerical results at finite temperature are obtained by using the Quantum Monte Carlo worm algorithm~\cite{Prokofev:1998} in the implementation of Ref.~\cite{Pollet:2007}. This is an unbiased and statistically exact path integral Monte Carlo algorithm, formulated in continuous imaginary time, in which two worm operators (corresponding to $\hat{a}$ and $\hat{a}^{\dagger}$ operators) perform local updates in an extended configuration space and hereby directly sample the Green function. These updates lead to a fast decorrelation between the configurations resulting in an integrated autocorrelation time of order unity. A bootstrap analysis shows that the relative error is below $10^{-3}$ for the presented data. It was previously demonstrated that experiments on the Bose-Hubbard model are in one-to-one agreement with such first principles simulations for realistic system sizes (on the order of a million atoms) and temperatures~\cite{Trotzky:2010}. In our simulations, we assume that temperature scales as $T \sim U$ which was shown to be a good assumption in this parameter regime~\cite{Pollet:2008}.

\end{document}